\documentclass[12pt]{article}
\usepackage{epsf}

\renewcommand{\bar}{\overline}
\newcommand{\newc}{\newcommand}
\newc{\shat}{\hat{s}}
\newc{\invpb}{\,\mbox{pb}^{-1}}
\newc{\pb}{\,\mbox{pb}}
\newc{\gev}{\,\mbox{GeV}}
\newc{\mev}{\,\mbox{MeV}}
\newc{\tev}{\,\mbox{TeV}}
\newc{\fract}{\frac}
\newc{\gsim}{\lower.7ex\hbox{$\;\stackrel{\textstyle>}{\sim}\;$}}
\newc{\lsim}{\lower.7ex\hbox{$\;\stackrel{\textstyle<}{\sim}\;$}}
\newc{\beq}{\begin{equation}}
\newc{\eeq}{\end{equation}}
\newc{\bea}{\begin{eqnarray}}
\newc{\eea}{\end{eqnarray}}
\newc{\eg}{{\it e.g.}}
\newc{\ie}{{\it i.e.}}
\newc{\etal}{{\it et al}}
\newc{\eps}{\epsilon}
\def\NPB#1#2#3{Nucl. Phys. {\bf B#1} (19#2) #3}
\def\PLB#1#2#3{Phys. Lett. {\bf B#1} (19#2) #3}

\def\PRD#1#2#3{Phys. Rev. {\bf D#1} (19#2) #3}

\def\ZPC#1#2#3{Zeit. f\"ur Physik {\bf C#1} (19#2) #3}

\catcode`@=11
\long\def\@caption#1[#2]#3{\par\addcontentsline{\csname
  ext@#1\endcsname}{#1}{\protect\numberline{\csname
  the#1\endcsname}{\ignorespaces #2}}\begingroup
    \small
    \@parboxrestore
    \@makecaption{\csname fnum@#1\endcsname}{\ignorespaces #3}\par
  \endgroup}
\catcode`@=12
\begin{document}

\begin{titlepage}
\begin{flushright}
{\rm
IASSNS-HEP-97-64\\
hep-ph/9705414\\
May 1997\\
}
\end{flushright}
\vskip 2cm
\begin{center}
{\Large\bf Implications of a charged-current anomaly
at HERA\footnote{Research
supported in part by DOE grant DE-FG02-90ER40542, by the W.~M.~Keck
Foundation, and by the Alfred P. Sloan Foundation.
Email: {\tt babu@sns.ias.edu, kolda@sns.ias.edu,
jmr@sns.ias.edu}}}
\vskip 1cm
{\large
K.S.~Babu,
Christopher Kolda\\
and\\
John March-Russell\\}
\vskip 0.4cm
{School of Natural Sciences\\ Institute for Advanced Study\\
Princeton, NJ 08540\\}
\end{center}
\vskip .5cm
\begin{abstract}
We demonstrate that in the presence of mixing between different
scalar leptoquark multiplets it is possible to simultaneously account
for the HERA high-$Q^2$ neutral current anomaly, and produce
a charged current anomaly of comparable magnitude.  The reduced
branching ratio to electrons and jets of the lightest leptoquark state 
results in a significant weakening of the CDF/D0 limits on scalar
leptoquarks; masses consistent with the HERA neutral current excess
are comfortably within the allowed range.  We show that the
possibilities for such a successful mixed leptoquark scenario are
quite limited, and we investigate some aspects of their phenomenology.
\end{abstract}
\end{titlepage}
\setcounter{footnote}{0}
\setcounter{page}{1}
\setcounter{section}{0}
\setcounter{subsection}{0}
\setcounter{subsubsection}{0}

%%%%%%%%%%%%%%%%%%%%%%%%%%%%%%%%%%%%%%%%%%%%%%%%%%%%%%%%%%%%%%%%%%%%%%%

\section{Introduction}

The H1~\cite{H1} and ZEUS~\cite{zeus} collaborations at HERA
recently announced an anomaly at high-$Q^2$ in the $e^+p\to eX$
neutral current (NC) channel.  With a combined luminosity of
$34.3\invpb$ in $e^+p\to eX$ mode at $\sqrt{s}=300\gev$, the two
experiments observe 24 events with $Q^2>15000\gev^2$ against
a Standard Model (SM) expectation of $13.4\pm1.0$, and 6 events
with $Q^2>25000\gev^2$ against an expectation of only $1.52\pm0.18$.
The high-$Q^2$ events seem to be clustered at Bjorken-$x$ values near
0.4 to 0.5, with the H1 data showing a more pronounced peak. 

HERA is capable of running in two modes: $e^-p$ and $e^+p$. In the former mode
H1 and ZEUS have accumulated $1.53\invpb$ of data but have observed
no statistically significant deviations from the SM. Further, the experiments
differentiate between final state $eX$ and $\nu X$, where the neutrino is
detected through its missing $p_T$.  H1 has also announced its
findings in the $e^+p\to\nu X$ charged current (CC) channel. They find
3 events at $Q^2>20000\gev^2$ with an expectation of $0.74\pm0.39$, but
no events with $Q^2>25000\gev^2$.  ZEUS has not announced its CC data.
Compared to the NC channel, the present CC signal is considerably
less statistically significant.  However, if the current trend
persists in the 1997 HERA data, there will be interesting constraints
on theoretical interpretations of the HERA excess.

There are three general categories of explanation for the NC excess: (1)
a statistical fluctuation, (2) unexpected SM physics, for example,
a modification of the parton distribution functions of the proton
at moderate-to-high-$x$ and large $Q^2$, or (3) new physics.

For case (1) we have nothing to say.  The possibility of modifying
the parton distribution functions (2) was suggested in~\cite{tung}, 
and its consequences for the high-$Q^2$ HERA NC and CC
data were studied in~\cite{bkmr}.  This suggestion has the advantage
that it is relatively conservative, and furthermore automatically
avoids the serious flavor problems~\cite{bkmrw,altarelli,limits}
associated with most
new physics explanations.  However the most attractive hypothesis
for modifying the parton distribution functions,
the (symmetric) intrinsic charm scenario, has already been shown
to be inconsistent with the high-$Q^2$ HERA data itself~\cite{bkmr}.

In category (3) are a variety of forms the new physics could
take.  These include contact interactions~\cite{altarelli,bkmrw,contact},
$s$-channel production of a leptoquark or a squark with $R$-parity
violating
interactions~\cite{early,rparity,altarelli,bkmrw,leptoquark,stirling},
and related proposals~\cite{others}. 

In this paper we consider the implications of the present
and future HERA CC data for new physics.  The constraints on, and the
consequences of, contact interactions for the CC data has been considered
by Altarelli, \etal~\cite{altarelliCC}; there they found that it was 
very difficult to explain any excess in the CC comparable to that in the NC.
They also considered CC signals in scenarios with leptoquarks, with 
emphasis on the supersymmetric $R$-parity violating case. 
For the case of leptoquarks they were able to find scenarios with
a significant CC excess. In this paper
we will also consider leptoquarks, but unlike Ref.~\cite{altarelliCC},
we will examine the consequences of leptoquark {\sl mixing} on the CC signal
and demonstrate that comparable CC and NC excesses can similarly be obtained. 
We also investigate the consequences of leptoquark mixing for low-energy
precision measurements, and for high-energy direct searches at the Tevatron.
In this regard, one noteworthy feature of scenarios with a CC signal is
that the recent CDF/D0 leptoquark limits~\cite{cdflq,d0lq,kramer}\ (for
scalar leptoquarks $M>210\gev$ at 95\% C.L.) are considerably weakened. 

If the NC anomaly at HERA is explained by $s$-channel production of a 
scalar leptoquark then the measured Bjorken-$x$ distribution 
translates into a mass determination: $m^2 =xs$.
For both H1 and ZEUS two independent
determinations are possible, depending on whether $x$ is calculated from
the double-angle or electron methods. For the seven events selected for
special study by H1 we find these two methods give $M_{2\alpha}=202\pm14\gev$
and $M_e=199\pm8\gev$ respectively. Similarly, for the five events
selected by ZEUS, $M_{2\alpha}=231\pm16\gev$ and $M_{e}=219\pm12\gev$.
Similarly, the 3 highest-$Q^2$ events in the H1 CC data are also clustered
in $x$ yielding $M_{CC}=197\pm20\gev$, where the error is dominated
by the H1 energy resolution. The fact that the CC and NC mass determinations
at H1 so closely match supports the hypothesis that both anomalies are
coming from production of the same particle, which then decays either 
to electrons or neutrinos.

\section{Mixing of scalar leptoquarks}

We will argue in this section that although with a single scalar
leptoquark multiplet
one does not expect a charged-current signal at HERA, in the presence
of mixing of different leptoquark multiplets a charged-current
signal is possible.

%%%%%%%%%%%%%%%%%%%%%%%%%
\begin{table}
\centering
\begin{tabular}{c|ccc|cc}
Operator & SU(3) & SU(2) & U(1) & CC & mode \\ \hline
$L^iQ^j\Phi'_{LQ}\eps_{ij}$ & $\bar3$ & 1 & ${1/3}$ & Yes & $e^-$ \\
$L^iQ^j(\Phi_{LQ})_{ij}$ & $\bar3$ & 3 & ${1/3}$ & Yes & $e^-$ \\
$L^i\bar u\Phi_{Lu}^j\eps_{ij}$ & 3 & 2 & ${7/6}$ & No & $e^+$ \\
$L^i\bar d\Phi_{Ld}^j\eps_{ij}$ & 3 & 2 & ${1/6}$ & No & $e^+$ \\
$\bar e Q^i\Phi_{eQ}^j\eps_{ij}$ & $\bar3$ & 2 & $-{7/6}$ & No & $e^+$ \\
$\bar e\bar u\Phi_{eu}$ & 3 & 1 & $-{1/3}$ & No & $e^-$ \\
$\bar e\bar d\Phi_{ed}$ & 3 & 1 & $-{4/3}$ & No & $e^-$
%\\ \hline
%$\bar L_i\gamma_\mu Q_j\Phi^\mu\eps^{ij}$ & $\bar3$ & 1 & $-\frac{2}{3}$
%& Yes & $e^+$ \\
%$\bar L_i\gamma_\mu Q_j\Phi^{\mu ij}$ & $\bar3$ & 3 & $-\frac{2}{3}$
%& Yes & $e^+$ \\
%$\bar L_i\gamma_\mu \bar u\Phi^\mu_j\eps^{ij}$ & 3 & 2 & $-\frac{1}{6}$
%& No & $e^-$ \\
%$\bar L_i\gamma_\mu\bar d\Phi^\mu_j\eps^{ij}$ & 3 & 2 & $\frac{7}{6}$
%& No & $e^-$ \\
%$e\gamma_\mu Q_i\Phi^\mu_j\eps^{ij}$ & $\bar3$ & 2 & $\frac{5}{6}$
%& No & $e^-$ \\
%$e\gamma_\mu\bar u\Phi^\mu$ & 3 & 1 & $\frac{5}{3}$ & No & $e^+$ \\
%$e\gamma_\mu\bar d\Phi^\mu$ & 3 & 1 & $\frac{2}{3}$ & No & $e^+$
\end{tabular}
\caption{List of scalar leptoquark operators. For each operator, the
$SU(3)\times SU(2)\times U(1)_Y$ quantum numbers of the leptoquark,
$\Phi$, are shown. The fifth column
indicates whether HERA should find CC events, and the final column lists the
mode in which HERA should dominantly produce the given leptoquark.
$Q$ and $L$ represent $SU(2)$ doublet quarks and leptons, while $\bar e$,
$\bar u$ and $\bar d$ are $SU(2)$ singlets.}
\label{optable}
\end{table}
%%%%%%%%%%%%%%%%%%%%%%%%%

The possible renormalizable couplings of scalar leptoquarks to
SM fermions are enumerated in Table~\ref{optable}~\cite{bkmrw}, along
with their $SU(3)\times SU(2)\times U(1)_Y$ quantum numbers.
We will consider only $s$-channel production of leptoquarks. 
(It was shown in~\cite{bkmrw} that $u$-channel exchange of a light leptoquark
could not consistently explain the HERA NC data.)  If the NC
events at HERA are due to an $s$-channel leptoquark, then the scattering
of the $e^+$ must
be off one of the valence quarks in the proton.\footnote{More specifically,
we are excluding scattering off of $\bar u$ or $\bar d$ quarks. 
We are not considering the possibility of scattering off of $s$ or $c$
sea quarks~\cite{rparity,altarelli,stirling}, where the required
Yukawa couplings must be of $O(1)$.}
Operators satisfying this constraint are listed as ``$e^+$'' mode in the 
last column of the table.  On the other hand, leptoquarks in the
remaining ``$e^-$'' operators are dominantly
produced in $e^-$ mode, and would have been already detected in HERA's
$e^-p$ data (even with the much smaller accumulated luminosity)
given the size of coupling necessary to explain the anomaly
in the $e^+p$ data.  The reason for this is that the leptoquarks in
``$e^-$'' operators are produced by scattering off sea quarks in $e^+p$
mode, but off valence quarks when HERA runs in $e^-p$ mode. 
Finally, in the column ``CC'' is indicated 
whether or not the operator leads, in the absence of mixing, to CC events
at HERA in addition to the NC.  We see that the three allowed types of 
scalar leptoquark are all SU(2) doublets, and cannot lead to a CC signal
at HERA in either $e^+ p$ or $e^-p$ mode.

However, as we now argue, in the presence of mixing, these three
operators can lead to a CC signal comparable in magnitude to the
NC signal.  
If we wish to simultaneously explain an excess in both the NC and CC channels,
then the mixing must involve at least one of the following leptoquark
states:
(i) the charge $Q=2/3$ component of $\Phi_{Ld}$, (ii) the $Q=5/3$
component of $\Phi_{Lu}$, and (iii) either the $Q=-5/3$ or the
$Q=-2/3$ components of $\Phi_{eQ}$. These components can 
mix with other leptoquark states or with each other.

The most important constraint on such mixing arises
from the necessity of avoiding helicity-unsuppressed rare decays.
If a leptoquark couples significantly to both left-handed and 
right-handed leptons and quarks,
it generates effective 4-Fermi operators of the
form $\bar{u_R}d_L\bar{e_R}\nu_L$.  These lead, for example, to
helicity-unsuppressed $\pi\to e\nu$ decays, which are severely
constrained by experiment~\cite{limits,mixing}. 
This eliminates the possibility of mixing $\Phi_{eQ}$ with any
leptoquark that would lead to a CC signal, since
by definition such a state has to couple to the left-handed lepton doublet 
containing a $\nu$.
Similarly, option (ii) is not realized, given that the only other $Q=5/3$
state is in $\Phi_{eQ}^*$.  Thus we are left with only option (i).

The $Q=2/3$ component of $\Phi_{Ld}$ can only mix with either
the $Q=2/3$ component of $\Phi_{Lu}$, or with the $Q=2/3$
component of the triplet $\Phi_{LQ}^*$.  Such mixings can arise from terms
in the Lagrangian, 
\beq
{\cal L} \supset \lambda\Phi_{Lu}^*\Phi_{Ld} H^2 + 
\tilde\mu\Phi_{Ld}\Phi_{LQ} H^*,
\label{eq:mixing}
\eeq
where $H$ is the SM Higgs doublet. In a supersymmetric context, we can
identify $H^*=H_d$ and $H=H_u$. The second term arises generically
in a softly-broken supersymmetric model, for instance from $A$-terms.
The first term seems to require additional fields. 

In Eq.~(\ref{eq:mixing}) the first term is $B$ and $L$ conserving, while the
second term violates $L$ (but not $B$) by $\Delta L=2$. Thus the second term
can lead to lepton number violating processes such as $\pi^+\to e^+\bar\nu_e$
which is not helicity-suppressed. The constraints on the $\Delta L=0$
process $\pi^+\to e^+\nu_e$ also apply here, requiring either leptoquark
masses over several TeV or Yukawa couplings of $\Phi_{LQ}$ too small to
be of use for producing a CC interaction at HERA. Thus we will not consider
this possibility any further. 

This leaves us with only the possibility of
$(\Phi_{Ld},\Phi_{Lu})$ mixing.
In principle {\sl both} the light and heavy mass eigenstate
admixtures of the $Q=2/3$ leptoquark states could be currently
produced by HERA.  This is because both have couplings to $e^+d$ after mixing,
and it is possible that they have comparable masses.  It is also 
possible to produce the $\phi_{Lu}^{5/3}$ state at HERA which leads to
no CC signal on its own. (We will return to
the constraints on such a scenario arising from CDF/D0 later.)
We think this is an interesting possibility, but we will focus
our discussion on the simpler possibility that only the lighter mixed
eigenstate is being produced at HERA; all the other states are
assumed to be heavy.  We denote this light mixed
state as $\phi_1^{2/3}=\cos\theta \phi_{Ld}^{2/3} +
\sin\theta \phi_{Lu}^{2/3}$, and the heavier mixed state (the orthogonal 
combination) as $\phi_2^{2/3}$.
The initial leptoquark Yukawa couplings before mixing are
\beq
{\cal L} = \lambda_1  L\bar{d} \Phi_{Ld} + \lambda_2 L\bar{u} \Phi_{Lu} + h.c.
\eeq
This leads to couplings of the light $\phi_1$ state of $(\lambda_1 \cos\theta
e\bar{d} + \lambda_2 \sin\theta \nu\bar{u})\phi_1^{2/3}$, and an
associated CC branching ratio
\beq
{\rm Br_{\nu j}}(\phi_1^{2/3} \to \bar{\nu}u) = {\lambda_2^2 \sin^2\theta \over
\lambda_1^2 \cos^2\theta + \lambda_2^2 \sin^2\theta}.
\label{Branch}
\eeq
For $\lambda_2 \sin\theta \sim \lambda_1\cos\theta$ this results in
comparable CC and NC signals at HERA.  Note that to maintain the overall
NC event rate, the effective coupling $\lambda_1\cos\theta$ (which was
$\simeq 0.04$ in the absence of mixing~\footnote{This includes the 
enhancement coming from NLO QCD effects~\cite{stirling,spira}.}) has to
be increased by a factor of $1/\sqrt{(1-{\rm Br_{\nu j}})}$.

It should be noted that CDF has recently announced
\cite{cdflq} a lower limit of $210\gev$ (at 95\% C.L.) on scalar
leptoquarks that decay into electrons and quarks with branching ratio
1, using the NLO QCD calculations of Ref.~\cite{kramer}. (The limits
on a leptoquark decaying to a quark and a neutrino are much weaker.)
This is starting to significantly constrain the (unmixed) leptoquark
explanation of the HERA NC data.  (From the calculations of
Ref.~\cite{blumleinvec}, it can be seen that this translates
into a limit on {\sl vector} leptoquarks that essentially excludes
this option for explaining the HERA NC anomaly.)

In the mixed scalar leptoquark scenario
the situation is significantly changed,
since the branching ratio to electrons and jets (${\rm Br}_{ej}=1-
{\rm Br}_{\nu j}$) is less than one.  For
example, if the branching ratio is 1/2, we estimate 
the mass limit on $\phi_1^{2/3}$
to be somewhat below $190 \gev$,
which is perfectly compatible with the HERA data.  In this case, however,
the heavier $\phi_2^{2/3}$ also decays into $e^+ + ~{\rm jets}$ with
a branching ratio of 1/2, so it contributes to any possible signal
at CDF/D0.  Setting the mass of $\phi_1^{2/3}$ at $200 \gev$, we estimate 
that the mass of $\phi_2^{2/3}$ must be $\gsim 225 \gev$ (for 
${\rm Br}_{ej}=1/2$) using the analysis of~\cite{kramer}.

One consequence of the CDF/D0 data is that it strongly disfavors
$\Phi_{eQ}$ as a possible explanation of the HERA NC anomaly.  
As we argued above, it is not possible to substantially mix
this state with any other which has couplings to neutrinos.  
Thus both components of $\Phi_{eQ}$ always decay into electrons +
jets.  Furthermore, since $\Phi_{eQ}$ is an unmixed doublet the
$\rho$-parameter constraint does not allow the mass of the heavier
component to be arbitrarily large.  Typically, one finds that the heavier
mass is constrained to be below $250\gev$, given that the lighter mass 
is fixed at $\simeq200\gev$.
Furthermore, there is a perturbativity constraint on 
the splitting since it is entirely due to electroweak symmetry breaking:
$\delta m^2\sim\lambda v^2\lsim(250\gev)^2$ for $\lambda\lsim1$.
Therefore CDF/D0 would see two states
each decaying to electrons with branching ratio one.  For instance
a heavy state at $250\gev$ contributes $0.03\pb$ to the CDF cross section,
which implies that the lighter state must be above $220\gev$ at 95\% C.L. 
Alternatively the quoted CDF limit for a single state implies a limit on  
a degenerate doublet of $M_{\Phi} > 234\gev$ at 95\% C.L., probably beyond
the interesting mass range for HERA.

\subsection{The $\rho$-parameter}

It is natural to consider the implications of the $\rho$-parameter 
constraints on scenarios in which scalar leptoquark multiplets mix. 
We have investigated this in detail following Ref.~\cite{einhorn}.  
In particular $\Delta \rho = (\Pi_{+-} - \Pi_{33})/m_W^2$ where the indices
label SU(2) weak-isospin, and the
general expression for the quantities $\Pi_{ab}$ is given by
\beq
\Pi_{ab} = -\frac{g^2C}{2}\int{d^4k_E\over (2\pi)^4} k^2_E
{\rm Tr}([T_a,\Delta(k)][T_b,\Delta(k)]).
\label{eq:piexpress}
\eeq
In this equation the $T$'s are SU(2) matrices in the (reducible)
representation of the scalar multiplets,
$\Delta(k)=1/(k_E^2 + M^2)$, with $M^2$ the mass squared matrix,
$g$ is the weak coupling, and $C$ is a possible color factor ($C=3$
for scalar leptoquarks).

The resulting analytic expressions for $\Delta \rho$ in the presence
of mixing between two SU(2) representations are quite complicated.  However,
in the case where two weak doublet leptoquarks (such as $\Phi_{Ld}$ and
$\Phi_{Lu}$) are mixed in such a way that
three of the four resulting leptoquarks are degenerate, the expressions
simplify greatly.  In this special case
\beq
\Delta \rho = {3G_F {\rm cos}^2 2\theta \over 8 \sqrt{2} \pi^2}\left[m_1^2 +
m_2^2+{2m_1^2m_2^2 \over m_1^2-m_2^2} {\rm ln}\left({m_1^2 \over m_2^2}\right)
\right]~,
\eeq
and it is clear then that for maximal mixing ($\theta=\pi/4$), $\Delta\rho$
vanishes.  For the general case we have investigated the contribution of
Eq.~(\ref{eq:piexpress})
to $\Delta \rho$ numerically.  In many cases it is possible to find substantial
{\sl negative} contributions to $\rho$.  For example, if we take the mass
eigenvalues to be $(202,273,240,240)\gev$, with maximal mixing between
the first two (the $Q=2/3$ states), 
then we find $\Delta \rho = -0.001$.  It is possible to 
find negative values of greater magnitude
(especially for large mixing), as well as small positive values, if one
allows the mass spectrum to vary consistent with the CDF/D0 limits.
Thus the $\rho$-parameter provides no significant constraint on the
mass spectrum in the $(\Phi_{Ld},\Phi_{Lu})$ mixed case.
(Note that if we require substantial mixing and one eigenvalue fixed around
$200\gev$, then perturbativity does not allow any of the other states to
become arbitrarily massive.)

\subsection{Atomic parity violation}

The mixing scenarios also have consequences for 
the atomic parity violation (APV) experiments.
In the presence of scalar leptoquarks the predicted weak
charge of a nucleus shifts by   
\beq
\Delta Q_W^{\rm LQ }=-2\left(\frac{\lambda_{LQ}/M_{LQ}}{g_W/M_W}\right)^2
(\delta_Z Z+\delta_N N).
\eeq
For the leptoquarks of interest, the values of $(\delta_Z,\delta_N)$
are: (2,1) for $\Phi_{Lu}$ and (1,2) for $\Phi_{Ld}$.

Recently the measurement of the weak charge $Q_W$ of Cesium
($Z=55$, $N=78$) has significantly improved~\cite{apv},
leading to a new value $Q_W^{\rm expt}({\rm Cs})=-72.11\pm0.93$
This is to be compared to the SM prediction of
$Q_W^{\rm SM}({\rm Cs})=-73.20\pm0.09$.
There are also reasonably good measurements for Thallium ($Z=81$, $N=205$),
leading to $Q_W^{\rm expt}({\rm Th})=-115.0\pm4.5$ compared to the
prediction $Q_W^{\rm SM}({\rm Th})=-116.8\pm0.19$~\cite{rosner}.
For the central values of the masses
and couplings of the $\Phi_{Lu}$ and $\Phi_{Ld}$ states found
in a fit to the NC HERA data ($\lambda_{Lu}\simeq 0.020$, $\lambda_{Ld}\simeq
0.04$, with masses $\simeq 200\gev$ in both cases), $Q_W({\rm Cs})$ shifts
by the small amounts $\Delta Q_W({\rm Cs})=-0.09$
and $\Delta Q_W({\rm Cs})=-0.40$, respectively, further away from 
experiment.

In the presence of mixing, the expression for $\Delta Q_W$ changes to
\beq
\Delta Q_W = -2{(\delta_Z Z +\delta_N N)\over (g_W/M_W)^2}\left[
\left({\lambda_1\cos\theta\over m_1}\right)^2 +
\left({\lambda_1\sin\theta\over m_2}\right)^2
\right] +\cdots
\label{eq:qwshift}
\eeq
The ellipsis denotes the contributions from the unmixed components
of the heavier multiplet, which depend upon its unknown Yukawa coupling
$\lambda_2$ to SM fermions. 
Since in the $(\Phi_{Ld},\Phi_{Lu})$ mixing case the combination
$\lambda_1\cos\theta$ is now fixed to be $\simeq 0.04$ by the magnitude of
the NC excess, it is clear that the $(\lambda_1\sin\theta/m_2)^2$ term
increases the shift in $\Delta Q_W$ compared to the unmixed case.  
For instance, in the case of maximal mixing one finds roughly a
doubling of $\Delta Q_W$, relative to the unmixed case.  (This 
shift is only increased by the terms proportional to $\lambda_2^2$
omitted in Eqn.~(\ref{eq:qwshift}).)
For Thallium, all the shifts are about twice as large,
but because of the much larger
experimental error this corresponds to a statistically less
significant change.  

Thus, overall, the discrepancy between
the APV measurements and the SM tends to increase in the case of 
significant $(\Phi_{Ld},\Phi_{Lu})$ mixing, by approximately
$0.75\sigma$ for Cesium (the exact value depending on the mass
$m_2$, the mixing angle, and the unknown Yukawa coupling $\lambda_2$).

\subsection{Heavy leptoquark decays}

We have seen from the discussions
above that the leptoquark interpretation of the HERA
NC data admits a comparably
large cross section in the charged current mode as well, provided
that the leptoquark produced is an admixture of
different leptoquark states.  The masses of these additional leptoquarks
are constrained by the $\rho$-parameter, and perturbativity, to be roughly
below $300\gev$.  It is then of interest to ask how
these extra leptoquarks could be detected once they are produced
at colliders.  Here we wish to point out some interesting
decay channels of these heavier leptoquarks which will be
of relevance for their search at the Tevatron.

For $(\Phi_{Ld},\Phi_{Lu})$ leptoquark mixing, since both are doublets, 
there are four physical states. The $\phi_{Ld}^{2/3}$
mixes with the $\phi_{Lu}^{2/3}$ forming two
mass eigenstates ($\phi_1^{2/3}$ and $\phi_2^{2/3}$) of
masses $m_1$ and $m_2$, with $m_1 \approx
200\gev$.  The two other states $\phi_{Ld}^{-1/3}$ and $\phi_{Lu}^{5/3}$ are
unmixed, and we denote their masses by $m_3$ and $m_4$.
The two body decay $\phi_{Ld}^{-1/3} \rightarrow d + \bar{\nu}$ is always
open, but the width for this decay is suppressed by
a small Yukawa coupling squared $\lambda^2 \sim (0.04)^2$.  If $m_3$ is larger
than $m_1+m_W$, the decay $\phi_{Ld}^{-1/3} \rightarrow \phi_1^{2/3}+W$ will
be open and will be dominant. One signature will be $3j+e$; a
more spectacular signature into $\ell\nu ej$ occurs about 1/3 as often.
If this channel is closed, there
is still the three-body decay through a virtual $W$ which can compete with
the $d + \bar{\nu}$ decay rate.  When the $W$ is off-shell,
the rate for the three-body decay is given by
\beq
\Gamma(\phi_{Ld}^{-1/3} \rightarrow \phi_1^{2/3}f\overline{f}) =
{A_s^2 A_f^2 \over 384 \pi^3} m_3 \int_0^{(1-\sqrt{r})^2}
dt {\left[r^2+(1-t)^2-2r(1+t)\right]^{3/2} \over
(t-s)^2}
\eeq
where $r \equiv m_1^2/m_3^2$, $s\equiv m_W^2/m_3^2$, $A_s = A_f=
g/\sqrt{2}$, and summation of different helicities is assumed.
The competing decay has a rate
\beq
\Gamma(\phi_{Ld}^{-1/3} \rightarrow
d+\bar{\nu})= {\lambda^2\over 16 \pi} m_3.
\eeq

Taking $m_1=200\gev$, the total width for the three body decay is found
to be (0.73 MeV, 2.0 MeV, 5.3 MeV) for $m_3= (250, 260, 270)\gev$.
These numbers are to be compared with the two-body decay
width $\Gamma\approx 7\mev$ for $\lambda = 0.04$.  
Identical results hold for the decay of the $\phi_{Lu}^{5/3}$
leptoquark.

What about the heavier $\phi_2^{2/3}$ leptoquark?  Since the mass
eigenstates are combinations of $Q=$2/3 states carrying
different $T_3$, there is a residual off-diagonal
$\phi_1^{2/3^*} \phi_2^{2/3} Z$ coupling with a coefficient
${[g \sin2\theta]/(2\cos\theta_W)}$, where
$\theta$ is the leptoquark mixing angle.  This leads to the three body
decay $\phi_2^{2/3} \rightarrow \phi_1^{2/3}+ Z^* \rightarrow
\phi_1^{2/3}+ f+\overline{f}$.  The rate is given by Eq. (1),
now with $A_s = g\sin2\theta/(2\cos\theta_W)$,
$A_f = g(T_3 -Q \sin^2\theta_W)/\cos\theta_W$, 
$r=m_1^2/m_2^2$ and
$s=m_Z^2/m_2^2$.  Again, this mode can compete with the 2-body decay mode.
For instance, if $m_2=(260,270,280)\gev$ with maximal mixing then the total
width for the 3-body decay is $(0.7, 1.8, 4.1)\,$MeV. If enough of the
heavier leptoquarks can be produced, the gold-plated
signature will be $\ell^+\ell^-e^+e^-jj$. Here the $\ell^+\ell^-$ come 
from the decay of $\phi_2$ to $Z^*\phi_1$ with the $\phi_1$ then further 
decaying to $ej$; the second $\phi_2$ on the other hand decays directly to
$ej$.

\section{Conclusions}

In the presence of mixing between different scalar leptoquark 
SU(2) multiplets we have demonstrated that it is possible to
simultaneously account for the HERA high-$Q^2$ NC anomaly, and produce
a CC anomaly of comparable magnitude.  The branching ratio to
electrons and jet(s) of the lightest leptoquark state 
that by assumption is being produced at HERA is consequently reduced.
This results in a significant weakening of the recently announced
CDF/D0 limits on scalar leptoquarks, and masses consistent with an
explanation of HERA are comfortably within the allowed range.

However we showed that there is only one possibility for such a successful 
mixed leptoquark scenario: the $\Phi_{Ld}$ doublet
leptoquark multiplet must mix with the $\Phi_{Lu}$ doublet.
We investigated various
aspects of the associated phenomenology.  In particular, this scenario
often allows relatively large splittings among the leptoquarks while
satisfying the $\rho$-parameter constraints.  Indeed it is possible
to find substantial negative contributions to $\rho$ in the case of
large mixing.  We also point out that there are novel possibilities
for the decays of heavier members of the leptoquark multiplets, which
might be relevant for searches at the Tevatron.

Finally, it should be emphasized that the mixed scenario does not address
what is in our opinion the most serious problem facing leptoquark
(or $R$-parity violating squark) explanations of the HERA anomaly, namely,
the very serious flavor problem that all such models
engender~\cite{bkmrw,altarelli,limits}.

\section*{Acknowledgments}

We wish to thank S.~Davidson, J.~Ellis, H.~Frisch, 
E.~Weinberg, and F.~Wilczek for useful conversations, and 
T.~Rizzo for a discussion on doublet-triplet leptoquark mixing.

\end{document}